\documentclass[
prl,
twocolumn,
floatfix,
superscriptaddress,
showpacs
]{revtex4}

\usepackage{graphicx}
\usepackage{epsfig}
\usepackage{latexsym}
\usepackage{amsmath}
\usepackage{amsfonts}
\usepackage{amsxtra}

\newcommand{\lwrsim}{\raise0.3ex\hbox{$<$\kern-0.75em\raise-1.1ex\hbox{$\sim$}}}
\def\krto{ {\,\,\lower .8ex\hbox {$\longrightarrow \atop k \rightarrow 0$}\,\,}}

\def\bea{\begin{eqnarray} }
\def\beq{\begin{eqnarray} }

\def\eea{\end{eqnarray}}
\def\eeq{\end{eqnarray}}

\def\eq#1{Eq.~(\ref{#1})}

\begin{document} 

\title{A novel method for the physical scale setting on the lattice and its application to $N_f$=4 simulations}

\author{(ETM Collaboration) B.~Blossier}
\author{Ph.~Boucaud} 
\affiliation{Laboratoire de Physique Th\'eorique, 
Universit\'e de Paris XI; B\^atiment 210, 91405 Orsay Cedex; France}
\author{M.~Brinet}
\affiliation{LPSC, CNRS/IN2P3/UJF; 
53, avenue des Martyrs, 38026 Grenoble, France}
\author{F.~De Soto}
\affiliation{Dpto. Sistemas F\'isicos, Qu\'imicos y Naturales, 
Univ. Pablo de Olavide, 41013 Sevilla, Spain}
\author{V.~Morenas}
\affiliation{Laboratoire de Physique Corpusculaire, Universit\'e Blaise Pascal, CNRS/IN2P3 
63177 Aubi\`ere Cedex, France}
\author{O.~P\`ene}
\affiliation{Laboratoire de Physique Th\'eorique, 
Universit\'e de Paris XI; B\^atiment 210, 91405 Orsay Cedex; France}
\author{K.~Petrov}
\affiliation{Laboratoire de
l'Acc\'el\'erateur Lin\'eaire, Centre Scientifique d'Orsay; B\^atiment 200, 91898 ORSAY Cedex, France}
\author{J.~Rodr\'{\i}guez-Quintero}
\affiliation{Dpto. F\'isica Aplicada, Fac. Ciencias Experimentales; 
Universidad de Huelva, 21071 Huelva; Spain.}
\affiliation{CAFPE, Universidad de Granada, E-18071 Granada, Spain}

\begin{abstract}

This letter reports on a new procedure for the lattice spacing setting that takes advantage of the 
very precise determination of the strong coupling in Taylor scheme. Although it can be applied 
for the physical scale setting with the experimental value of $\Lambda_{\overline{\rm MS}}$ as 
an input, the procedure is particularly appropriate for relative ``calibrations". 
The method is here applied for simulations with four degenerate light quarks in the sea and 
leads to prove that their physical scale is compatible with the same one for simulations with two 
light and two heavy flavours.

\begin{flushright}
LPT-Orsay 13-92\\
UHU-FT/13-11 \\
\end{flushright}

\end{abstract}


\maketitle



\section{Introduction}

The field theory of the strong interactions, QCD, is essentially nonperturbative in its low energy domain. 
There, its asymptotic states differ from the non-interacting elementary fields and it should properly account for 
the main features of the strong phenomenology: chiral symmetry breaking and confinement. 
One of the most fruitful nonperturbative approaches to the QCD low-energy phenomenology is 
the lattice field theory~\cite{Wilson:1974sk} which, more and more in the last few years, is providing with accurate 
numerical results to account for the rich phenomenology of QCD~\footnote{See the lattice review of \cite{Beringer:1900zz} or 
the contributions in PoS(Lattice 2012).}. To this goal, a key role is played by 
the physical scale setting or lattice ``calibration": the adjustement of the lattice spacing to reproduce properly 
a low-energy experimental value: masses, decay constants, etc. 

The purpose of this letter is to present a novel technique to perform this scale setting, which is based on the 
direct computation of the strong coupling constant from the gauge and ghost propagators. In the past, gluonic 
quantities, as the string tension for the linear static interquark potential~\cite{Bali:1992ab,Bali:1992ru,Sommer:1993ce,Boucaud:1998bq}, has been used to perform a relative 
calibration: to fix the lattice spacing for one simulation from that known from another different simulation. 
The method presented here avails for a relative calibration from gluonic quantities but, the strong coupling 
being directly accessible from experiments, also for an absolute lattice calibration with $\Lambda_{\rm QCD}$ as an input. 
This is particularly useful for simulations with many degenerate light flavours, as those to compute 
renormalization constants in the flavour massless limit~\cite{ETM:2011aa} or motivated by the expected similarities of many-light-flavours QCD with Walking models for technicolor~\cite{Holdom:1984sk} as Refs.~\cite{Appelquist:2009ka,Iwasaki:2012kv,Aoki:2013xza,Aoki:2013dz,Bashir:2013zha}. In those cases, there is no physical quantity to compare with for the scale setting, but $\Lambda_{\rm QCD}$ can be well defined 
by assuming the strong coupling running not to depend on the quark masses, at least far away from 
the flavour thresholds. Furthermore, for more than 2 light degenerate flavours and 3 Goldstone 
bosons, the standard chiral behaviour cannot be reliably applied to guide the chiral fits of 
masses or decay constants. On the other hand, $\Lambda_{\rm QCD}$ being the fundamental scale of 
QCD, to which many different experiments refer, to use it for the scale setting could be taken as 
a theoretical ``{\it ace}". Last but not least, the strong coupling running being 
obtained from data for different simulations, the results can be compared to each other and directly confronted 
to analytical QCD predictions. This provides with a very valuable crosscheck for the scale setting reliability and 
ensures the best accuracy.

\section{The matching by the Taylor coupling}

The strategy is to get the ratios of lattice spacings from different simulations 
by the intercomparison of a renormalization-group invariant (RGI) quantity, as 
the one defining a coupling, computed with the lattice gauge field configurations obtained 
from the simulations. 

Let's call $Q$ this quantity that could be computed from lattice QCD such that one 
would have 
\beq\label{eq:Qlatt}
Q_{\rm phys}\left( p \right) \ = \  Q_{\rm Latt}(p_L)  + {\cal O}(a) \ ,
\eeq
where the physical and the lattice momenta are related such that
$p_L=a(\beta,\mu) p$, where $a(\beta,\mu)$ stands for the lattice spacing. 
We consider a particular simulation in a $N^3\times N_t$ lattice 
with $\beta$, for the bare lattice coupling, and $\mu$, standing for any other 
relevant set-up parameter (in our next application, the twisted mass of the light~\footnote{The dependence in the heavy masses will be dealt with at length in this paper.} degenerated 
quarks~\cite{Frezzotti:2000nk,Frezzotti:2003xj}).
After the appropriate Fourier transform of data in configuration space from the simulations, 
one would be left with
\beq\label{eq:pL}
p_L^2 = \left( \frac{2 \pi}{N} \right)^2 \left( n_x^2 + n_y^2 + n_z^2 + \frac{N^2}{N_t^2} n_t^2 \right) \ ,
\eeq
defined by the four integers $n_x$, $n_y$, $n_z$ and $n_t$. In \eq{eq:Qlatt}'s r.h.s., we included terms of the order 
$a$ to account for the lattice artefacts that should tend to disappear when approaching the continuum limit. 
$Q_{\rm phys}$ will be now supposed not to depend on the lattice set-up parameters at sufficiently high energy 
where the matching is possible, such that, 
for two different simulations with parameters $(\beta_1,\mu_1)$ and 
$(\beta_2,\mu_2)$, after neglecting (or properly correcting) the lattice artefacts, we can write: 
\beq\label{eq:match}
Q_{\rm Latt}^{(\beta_1,\mu_1)}(p_L) \ \equiv \   Q_{\rm phys}\left(p\right) 
\equiv \ 
Q_{\rm Latt}^{(\beta_2,\mu_2)}\left( p'_L \right)  , 
\eeq
where:  $ p'_L / a(\beta_2,\mu_2) = p_L / a(\beta_1,\mu_1) = p$. 
Then, the ratio of lattice spacings, $a(\beta_2,\mu_2)/a(\beta_1,\mu_1)$ is to be obtained by 
computing $Q$ from the two different simulations and impose the results to match as \eq{eq:match} requires. 
The latter implies that, where 
the matching is required, any dependence of $Q$ on $\beta$ and $\mu$ has been supposed to be captured 
by the lattice spacing through the scale setting. This will be confirmed, in our procedure, by the comparison 
of the running of $Q$ with the momentum for the different simulations, after the scale setting. 

In order to apply the matching procedure above described, we will choose for $Q$ the Taylor coupling~\cite{vonSmekal:1997is,Boucaud:2008gn,Aguilar:2009nf,Aguilar:2010gm}, 
\beq\label{eq:QLatt}
Q_{\rm Latt}(p_L) \equiv \alpha_T^{\rm Latt}(p_L) = \frac{g^2_0(a)}{4 \pi} \widetilde{Z}_3^2(p_L,a) Z_3(p_L,a) \ ,
\eeq
estimated  from different lattice simulations, where $\widetilde{Z}_3$ and $Z_3$ are the ghost and gluon propagator 
renormalization constants in MOM scheme.  
Then, we will first apply the so-called $H(4)$-extrapolation procedure~\cite{Becirevic:1999uc,Becirevic:1999hj,deSoto:2007ht}, 
that exploits the remaining symmetry which is restricted to the $H(4)$ isometry group for 
the elimination of $O(4)$-breaking lattice artefacts, 
\beq\label{eq:H4}
\alpha_T^{\rm Latt}(p_L) \ \Rightarrow 
\alpha_T^{H(4)}(p_L) \ .
\eeq
We will next correct for 
the $O(4)$-invariant lattice artefacts as shown in Refs.~\cite{Blossier:2011tf,Blossier:2012ef},
\beq\label{eq:alphaLatt}
\alpha_T^{H(4)}(p_L) \ = \ \alpha_T^{\rm phys}\left( a^{-1} p_L \right) \ + \ c_{a^2p^2} \ p_L^2 \ ,
\eeq 
and will finally be left with the ``{\it continuum}" Taylor coupling that have been shown to be, 
in practice, very well described by
\beq\label{eq:alphahdp6}
\alpha_T^{\rm phys}(p) \ = \ \alpha_T(p^2) \ + \ \frac{d_x}{p^x} \ ,
\eeq
with
\beq\label{alphahNP}
\alpha_T(p^2)
&=& 
\alpha^{\rm pert}_T(p^2)
\ 
\left( \rule[0cm]{0cm}{0.85cm}
 1 + \frac{9}{p^2} \
R\left(\alpha^{\rm pert}_T(p^2),\alpha^{\rm pert}_T(q_0^2) \right) \right.
\nonumber \\
&\times& \left. \left( \frac{\alpha^{\rm pert}_T(p^2)}{\alpha^{\rm pert}_T(q_0^2)}
\right)^{1-\gamma_0^{A^2}/\beta_0} 
\frac{g^2_T(q_0^2) \langle A^2 \rangle_{R,q_0^2}} {4 (N_C^2-1)}
\right) , 
\eeq
derived from the OPE description of ghost and gluon dressing functions in terms of the dimension-two gluon 
condensate, $R(\alpha,\alpha_0)$ including higher-order corrections to the Wilson 
coefficient beyond the leading logarithm.
The purely perturbative running is given by $\alpha_T^{\rm pert}$ up to four-loops through the integration 
of the $\beta$-function~\cite{Beringer:1900zz},  
in terms of $\ln(p/\Lambda_T)$ where $\Lambda_T/\Lambda_{\overline{\rm MS}}=0.5608$ for $N_f$=4. 
In ref.~\cite{Blossier:2012ef}, Eqs.~(\ref{eq:alphaLatt}-\ref{alphahNP}) have been successfully applied 
to fit the running of the Taylor coupling obtained from unquenched lattice simulations with two 
light degenerate quark flavours and two heavier non-degenerate ones ($N_f$=2+1+1). The results, 
recently updated in ref.~\cite{Blossier:2013nw}, for the best-fit parameters are: $\Lambda_{\overline{\rm MS}} \overline{a}(1.90,0)=0.1413(32)$, $g^2\langle A^2 \rangle \overline{a}^2(1.90,0)=0.76(11)$, $x=5.73(27)$ and $d_x \overline{a}^x(1.90,0)=-0.157(10)$; 
expressed in units of $\overline{a}(1.90,0)$ (we used here $\overline{a}$ for 
simulations with $N_f$=2+1+1 and keep $a$ for $N_f$=4 degenerate flavours). 

Then, for any simulation with set-up parameters $(\beta,\mu)$, according 
to Eqs.~(\ref{eq:match}-\ref{eq:alphaLatt}), one can write
\beq\label{eq:fit}
\alpha_{T,(\beta,\mu)}^{H(4)}\left(p_L \right) &=& 
\alpha^{\rm phys}_T\left(\frac{p'_L}{\overline{a}(1.90,0)}\right) 
+ c_{a^2p^2} \ p_L^2 \ ,
\eeq
where $p'_L=p_L \overline{a}(1.90,0)/a(\beta,\mu)$, $p_L$ being the lattice momentum for the simulation, \eq{eq:pL}, 
and the running of $\alpha_T^{\rm phys}$ given by Eqs.~(\ref{eq:alphahdp6}-\ref{alphahNP}) and 
expressed in units of $\overline{a}(1.90,0)$, 
\beq\label{eq:fit2}
\alpha_T^{\rm phys}\left(\frac{p'_L}{\overline{a}(1.90,0)}\right) &=& \nonumber \\ 
\alpha_T\left(\frac{{p'_L}^2}{\overline{a}^2(1.90,0)}\right) &+& \frac{d_x \overline{a}^x(1.90,0)}{{p'_L}^x} , 
\eeq
with $x=5.73$ and the central value for the parameters $\Lambda_{\overline{\rm MS}} \overline{a}(1.90,0)$, 
$g^2\langle A^2 \rangle \overline{a}^2(1.90,0))$ and  $d_x \overline{a}^x(1.90,0)$ above presented. 
The latter is a consequence of our main assumption: $\Lambda_{\overline{\rm MS}}$ and the nonperturbative corrections, 
coded by $g^2\langle A^2 \rangle$ and $d_x$, are supposed to depend only on the number of active quarks 
and, far above the quark mass thresholds, their masses should not matter so much. As the matching 
of coupling data for simulations with $\mu$ and $\mu'$ set-up parameters naturally implies~\footnote{Deviations from \eq{eq:scaling} should be included in the procedure's systematic uncertainties.}
\beq\label{eq:scaling}
\frac{\Lambda_{\overline{\rm MS}}^{\mu'}}{\Lambda_{\overline{\rm MS}}^{\mu}} 
\simeq \left(\frac{g^2\langle A^2 \rangle^{\mu'}}{g^2\langle A^2 \rangle^{\mu}}\right)^{1/2} 
\simeq \left(\frac{d_x^{\mu'}}{d_x^{\mu}}\right)^{1/x} \simeq 1 \ ,
\eeq
this main assumption will appear supported ``{\it a posteriori}" (see next Fig.~\ref{fig:matching}). 
Thus, taking the ratios in \eq{eq:scaling} to be exactly 1, the ratio of lattice spacings, $\overline{a}(1.90,0)/a(\beta,\mu)$, and the coefficient $c_{a^2p^2}$ are the two only free parameters to be determined by the best fit of Eqs.~(\ref{eq:fit},\ref{eq:fit2}) to the Taylor coupling lattice data.

\section{Relative calibration}

In the following, the above-described procedure will be applied to estimate the lattice spacing 
for simulations with $N_f$=4 degenerate twisted-mass flavours~\cite{ETM:2011aa} (Tab.~\ref{tab:param} gathers their set-up parameters), produced by ETMC to apply the massless renormalization.
To our knowledge, no other method allows for such a reliable scale setting in this case, 
as the Taylor coupling can be properly taken not to depend very much~\footnote{This is the case, as the matching we reach 
shows, at least for quark masses varying not too much, as happens for our simulations.} on the set-up parameters for 
$N_f$=4 and $N_f$=2+1+1 simulations.  

\begin{table}[h]
\begin{center}
\begin{tabular}{|c|c|c|c|c|}
\hline
\hline
$\beta$ & $a \mu$ & $a m_{\rm PCAC}$ & $a M_0$ & confs. \\
\hline
\hline
1.90 & 0.0080 & -0.0390(01) & 0.0285(01) & 130 \\
1.90 & 0.0080 & 0.0398(01) & 0.0290(01) & 130 \\
\hline
1.90 & 0.0080 & -0.0358(02) & 0.0263(01) & 200 \\
1.90 & 0.0080 & 0.0356(01) & 0.0262(01) & 200 \\
\hline
1.90 & 0.0080 & -0.0318(01) & 0.0237(01) & 200 \\
1.90 & 0.0080 & 0.0310(02) & 0.0231(01) & 200 \\
\hline
1.90 & 0.0080 & -0.0273(02) & 0.0207(01) & 130 \\
1.90 & 0.0080 & 0.0275(04) & 0.0209(01) & 130 \\
\hline
\hline
1.95 & 0.0085 & -0.0413(02) & 0.0329(01) & 130 \\
1.95 & 0.0085 & 0.0425(02) & 0.0338(01) & 130 \\
\hline
1.95 & 0.0085 & -0.0353(01) & 0.0285(01) & 130 \\
1.95 & 0.0085 & 0.0361(01) & 0.0285(01) & 130 \\
\hline
1.95 & 0.0020 & -0.0363(01) & 0.0280(01) & 120 \\
1.95 & 0.0020 & 0.0363(01) & 0.0274(01) & 120 \\
\hline
1.95 & 0.0180 & -0.0160(02) & 0.0218(01) & 130 \\
1.95 & 0.0180 & 0.0163(02) & 0.0219(01) & 130 \\
\hline
1.95 & 0.0085 & -0.0209(02) & 0.0182(01) & 130 \\
1.95 & 0.0085 & 0.0191(02) & 0.0170(01) & 130 \\
\hline
1.95 & 0.0085 & -0.0146(02) & 0.0141(01) & 130 \\
1.95 & 0.0085 & 0.0151(02) & 0.0144(01) & 130 \\
\hline
\hline
2.10 & 0.0078 & -0.00821(11) & 0.0102(01) & 180 \\
2.10 & 0.0078 & 0.00823(08) & 0.0102(01) & 180 \\
\hline
2.10 & 0.0064 & -0.000682(13) & 0.0084(01) & 180 \\
2.10 & 0.0064 & 0.00685(12) & 0.0084(01) & 180 \\
\hline
2.10 & 0.0046 & -0.00585(08) & 0.0066(01) & 120 \\
2.10 & 0.0046 & 0.00559(14) & 0.0064(01) & 120 \\
\hline
2.10 & 0.0030 & -0.00403(14) & 0.0044(01) & 240 \\
2.10 & 0.0030 & 0.00421(13) & 0.0045(01) & 240 \\
\hline
\hline
\end{tabular}
\end{center}
\caption{Set-up parameters, $m_{\rm PCAC}$ and the bare polar mass for the ensembles here exploited. Borrowed from 
ref.~\cite{ETM:2011aa}.}
\label{tab:param}
\end{table}

\begin{figure}
\begin{center}
\includegraphics[width=0.90\columnwidth]{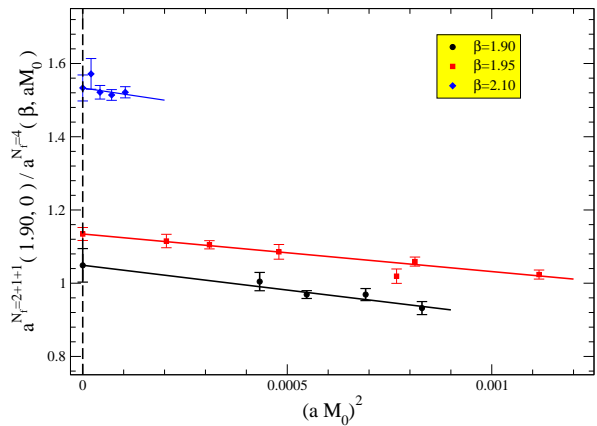}
\end{center}
\vspace*{-0.75cm}
\caption{\small ratios of lattice spacings (see tab.~\ref{tab:ratios}) obtained 
by the matching procedure of the Taylor coupling and the corresponding chiral extrapolation 
in solid line.}
\label{fig:ratios}
\end{figure}

We will take the lattice spacing to depend on the bare gauge coupling, $\beta$, and on the 
dynamical degenerate-flavour mass only through the bare polar mass, $M_0$ (see Tab.~\ref{tab:param} and 
Ref. \cite{ETM:2011aa}). Then, we compute the Taylor coupling, $\alpha_T^{\rm Latt}$, given by \eq{eq:QLatt} 
for each lattice ensemble. Next, we average for the two ensembles with roughly the same $m_{\rm PCAC}$ but opposite sign, 
as explained in ref.~\cite{ETM:2011aa}, in order to achieve approximatively the $O(a)$ improvement 
though working out of the maximal twist. We apply the $H4$ extrapolation procedure 
to remove the hypercubic artefacts and, finally, the cured results for the coupling 
is fitted with  Eqs.~(\ref{eq:fit},\ref{eq:fit2}), as explained in the previous section.
Thus, we obtain the ratios of lattice spacings, $\overline{a}(1.90,0)/a(\beta,aM_0)$, 
and $c_{a^2p^2}$ as the best-fit parameters gathered in Tab.~\ref{tab:ratios}. 
The results appear also plotted in Fig.~\ref{fig:ratios}, where a linear extrapolation on $M_0^2$, as the 
use of a ${\cal O}(a)$-improved lattice action suggests, down to the chiral limit is also shown.
 It should be noticed that the fitted parameters for the coefficient correcting the $O(4)$-invariant lattice 
artefacts, $c_{a^2 p^2}$, shows no important dependence on the light quark mass, as expected, and fairly 
well agree with the same parameter obtained for our previous analysis with simulations for 
$N_f=$2+1+1~\cite{Blossier:2012ef,Blossier:2013nw}. 

\begin{table}[h]
\begin{center}
\begin{tabular}{|c|c|c|c|c|}
\hline
$\beta$ & $aM_0$ & $\overline{a}(1.90,0)/a(\beta,aM_0)$ & $c_{a^2p^2}$ & $\chi^2$/d.o.f.\\
\hline
\hline
1.90 & 0.0288 & 0.932(18) & -0.0074(14) & 5.6/44 \\ 
\hline
1.90 & 0.0263 & 0.969(17) & -0.0067(5) & 3.3/45 \\ 
\hline
1.90 & 0.0234 & 0.969(11) & -0.0080(10) & 7.6/45 \\
\hline
1.90 & 0.0208 & 1.004(25) & -0.0056(12) & 2.4/46 \\
\hline
\hline
1.90 & 0 & 1.049(46) & & \\
\hline
\hline
1.95 & 0.0334 & 1.024(12) & -0.0079(8) & 4.4/46 \\ 
\hline
1.95 & 0.0285 & 1.059(12) & -0.0088(9) & 10.2/49 \\ 
\hline
1.95 & 0.0277 & 1.019(20) & -0.0099(9) & 5.6/46 \\ 
\hline
1.95 & 0.0219 & 1.086(20) & -0.0069(9) & 3.4/47 \\
\hline
1.95 & 0.0176 & 1.105(11) & -0.0066(11) & 19.8/48 \\
\hline
1.95 & 0.0143 & 1.115(18) & -0.0054(7) & 9.9/50 \\
\hline
\hline
1.95 & 0 & 1.134(18) & & \\
\hline
\hline
2.10 & 0.0102 & 1.530(15) & -0.0053(4) & 142./93\\ 
\hline
2.10 & 0.0084 & 1.518(15) & -0.0048(5) & 90.3/93\\ 
\hline
2.10 & 0.0065 & 1.533(19) & -0.0049(5) & 161./93\\
\hline
2.10 & 0.0045 & 1.578(42) & -0.0055(10) & 240./94 \\
\hline
\hline

2.10 & 0 & 1.533(35)  & &
\\
\hline
\hline
\end{tabular}
\end{center}
\caption{\small ratios of lattice spacings obtained as explained in the text. The values obtained by performing a chiral extrapolation down to a zero light quark mass is also shown. The quality of the fits is characterized by 
the $\chi^2$/d.o.f. All the errors have been derived by applying the Jacknife's method.}
\label{tab:ratios}
\end{table}


In Tab.~\ref{tab:comp}, the ratios of $N_f$=4 lattice spacings over that at $\beta=1.90$ for $N_f$=2+1+1  
from Tab.~\ref{tab:ratios}, after the chiral extrapolation, are shown in comparison with ratios of the same 
lattice spacings for $N_f$=2+1+1, borrowed from Refs.~\cite{Blossier:2013nw,Silvano:2013nw}. They all agree 
within the errors, although the lattice spacings for $N_f$=4 appear to be systematically larger ($\sim 5$ \%) than those 
for $N_f$=2+1+1. 

\begin{table}[h]
\begin{center}
\begin{tabular}{|c|c|c|c|}
\hline
$\beta$ & $\overline{a}(1.90,0)/a(\beta,0)$ & $a(1.90,0)/a(\beta,0)$ & $\overline{a}(1.90,0)/\overline{a}(\beta,0)$ \\
\hline
\hline
1.90 & 1.049(46) & 1 & 1  \\
\hline
1.95 & 1.134(18) & 1.081(50) & \cite{Silvano:2013nw}:~1.085(59) \\
\hline
2.10 & 1.533(35) & 1.461(72) & 
\begin{tabular}{c} \cite{Blossier:2013nw}:~1.477(28)  
\\
\cite{Silvano:2013nw}:~1.429(71)
\end{tabular}
\\
\hline
\hline
\end{tabular}
\end{center}
\caption{\small Comparison of the ratios of lattice spacings for $N_f$=4 (noted as $a$) 
obtained here and those for $N_f$=2+1+1 simulations (noted as $\overline{a}$) from Refs.~\cite{Blossier:2013nw,Silvano:2013nw}.}
\label{tab:comp}
\end{table}


\section{Absolute calibration from $\Lambda_{\overline{\rm MS}}$}

In the previous section, the matching of the Taylor coupling led to a relative scale setting for the 
analysed simulations, {\it i.e.} in terms of a given lattice spacing for another simulation 
($\beta=1.90$ and $N_f$=2+1+1, with chiral light flavours). 
Then, the ``absolute" calibration of the former, in physical units,  
requires from the latter's knowledge. 
On the other hand, the Taylor coupling from lattice data confronted to Eqs.~(\ref{eq:alphaLatt},\ref{alphahNP}) 
provided with an estimate for $\Lambda_{\overline{\rm MS}}$ in terms of the lattice spacing. Such an estimate 
was used in Refs.~\cite{Blossier:2011tf,Blossier:2012ef,Blossier:2013nw} to compute, 
after the scale setting from ETMC, $\Lambda_{\overline{\rm MS}}$ in physical units and 
hence $\alpha_{\overline{\rm MS}}(m^2_Z)$. Alternatively, one can also take 
the experimental value for $\Lambda_{\overline{\rm MS}}$ and use it to estimate the lattice spacing. 
We have $\overline{a}(1.90,0) \Lambda_{\overline{\rm MS}}=0.1413(32)$, from lattice data 
with $N_f$=2+1+1 unquenched flavours, as mentioned above, and 
$\Lambda_{\overline{\rm MS}}^{N_f=4}=296(10)$ MeV 
from PDG~\cite{Beringer:1900zz}. Then, for $N_f$=2+1+1, one would have
$\overline{a}(1.90,0)=0.0940(38)~{\rm fm}$, which compares fairly well to the very 
recent ETMC result: $\overline{a}(1.90,0)=0.0885(36)~{\rm fm}$~\cite{Silvano:2013nw}.
It should be furthermore noticed that the determination of $\overline{a}(1.90,0) \Lambda_{\overline{\rm MS}}$ in ref.~\cite{Blossier:2013nw} takes into account systematic uncertainties we do not include in the present calibration. 
These uncertainties could be drastically reduced by performing a simulation at as larger a $\beta$ parameter as possible, 
to reach larger physical momenta but keeping the higher-order hypercubic artefacts under control.

\begin{figure}
\begin{center}
\includegraphics[width=0.90\columnwidth]{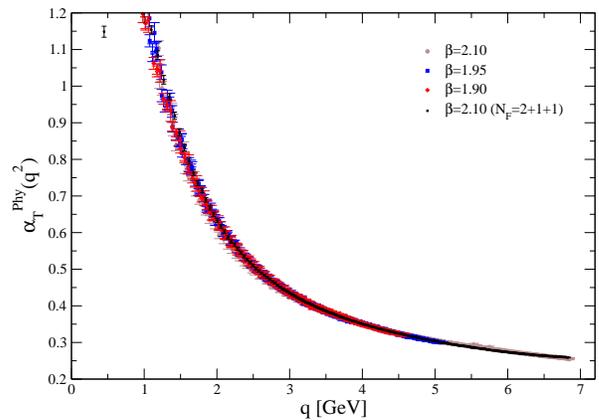}
\end{center}
\vspace*{-0.75cm}
\caption{\small The physical running of the Taylor coupling, defined by \eq{eq:alphaLatt}, for all the properly calibrated 
lattices from Tab.~\ref{tab:param}. $N_f=$2+1+1 data from~\cite{Blossier:2013nw} are included for comparison.}
\label{fig:matching}
\end{figure}

Thus, we can take our estimate for the lattice spacing and 
set the physical scale for all the lattices in Tab.~\ref{tab:param}, with the help of the 
ratios from Tab.~\ref{tab:ratios}. Then, we verify that the running of $\alpha_T^{\rm phys}$, 
defined by \eq{eq:alphaLatt}, is the same for all of them and the same as for $N_f$=2+1+1, as can be seen in Fig.~\ref{fig:matching}. In particular, in the chiral limit, we obtain
\beq\label{eq:abs2}
\frac{a(\beta,0)}{1 \mbox{\rm fm}} =
\left\{ 
\begin{tabular}{lr}
0.0896(53) & \ \ $\beta=1.90$ \\
0.0829(36) & \ \ $\beta=1.95$ \\
0.0613(29) & \ \ $\beta=2.10$ 
\end{tabular}
\right. \ ,
\eeq
for the $N_f=4$ simulations. 

\section{Conclusions}

We have proposed a novel method for the scale setting on lattice simulations that only needs the 
evaluation of gauge and ghost propagators to determine the strong coupling running and requires for 
it, after the appropriate removal of lattice artefacts, to be the same for different simulations, 
when the scale is properly fixed. The method allows for a relative calibration of lattices, 
the lattice spacing for them being expressed in terms of the one in another given simulation, but 
also for an absolute calibration with $\Lambda_{\overline{\rm MS}}$ as an input. The method 
has been successfully applied to perform the scale setting for unquenched simulations including 
four degenerate light flavours. Thus, we have also concluded that, within our statistical uncertainties, 
the lattice spacings for $N_f$=4 and $N_f$=2+1+1 simulations appear to be compatible.



\section*{Acknowledgements} 

We thank the support of Spanish MICINN FPA2011-23781 and  
the HPC resources from CC-IN2P3 (CNRS-Lyon), IDRIS (CNRS-Orsay),   
GENCI (Grant 052271).


\bibliography{total}

\end{document}